\documentclass[aip,rsi,amsmath,amssymb,reprint,superscriptaddress]{revtex4-1}
\usepackage{graphicx}
\usepackage{dcolumn}
\usepackage{bm}
\begin{document}
\title{Broadband, large-area microwave antenna for optically-detected magnetic resonance of nitrogen-vacancy centers in diamond}
\author{Kento Sasaki}
\affiliation{School of Fundamental Science and Technology, Keio University, 3-14-1 Hiyoshi, Kohoku-ku, Yokohama 223-8522, Japan}
\author{Yasuaki Monnai}
\affiliation{School of Fundamental Science and Technology, Keio University, 3-14-1 Hiyoshi, Kohoku-ku, Yokohama 223-8522, Japan}
\author{Soya Saijo}
\affiliation{School of Fundamental Science and Technology, Keio University, 3-14-1 Hiyoshi, Kohoku-ku, Yokohama 223-8522, Japan}
\author{Ryushiro Fujita}
\affiliation{School of Fundamental Science and Technology, Keio University, 3-14-1 Hiyoshi, Kohoku-ku, Yokohama 223-8522, Japan}
\author{Hideyuki Watanabe}
\affiliation{Correlated Electronics Group, Electronics and Photonics Research Institute, National Institute of Advanced Industrial Science and Technology (AIST),
Tsukuba Central 5, 1-1-1, Higashi, Tsukuba, Ibaraki 305-8565, Japan}
\author{Junko Ishi-Hayase}
\affiliation{School of Fundamental Science and Technology, Keio University, 3-14-1 Hiyoshi, Kohoku-ku, Yokohama 223-8522, Japan}
\author{Kohei M. Itoh}
\email{kitoh@appi.keio.ac.jp}
\affiliation{School of Fundamental Science and Technology, Keio University, 3-14-1 Hiyoshi, Kohoku-ku, Yokohama 223-8522, Japan}
\author{Eisuke Abe}
\email{e-abe@keio.jp}
\affiliation{School of Fundamental Science and Technology, Keio University, 3-14-1 Hiyoshi, Kohoku-ku, Yokohama 223-8522, Japan}
\date{\today}

\begin{abstract}
We report on a microwave planar ring antenna specifically designed for optically-detected magnetic resonance (ODMR) of nitrogen-vacancy (NV) centers in diamond.
It has the resonance frequency at around 2.87~GHz with the bandwidth of 400~MHz,
ensuring that ODMR can be observed under external magnetic fields up to 100~G without the need of adjustment of the resonance frequency.
It is also spatially uniform within the 1-mm-diameter center hole,
enabling the magnetic-field imaging in the wide spatial range.
These features facilitate the experiments on quantum sensing and imaging using NV centers at room temperature.
\end{abstract}
\maketitle

\section{Introduction\label{intro}}
Nitrogen-vacancy (NV) center in diamond is an atomic defect consisting of a substitutional nitrogen and a vacancy adjacent to it.
Among a few charge states NV centers can take, the negatively-charged NV$^-$ state possesses optical and magnetic properties favorable to quantum technology applications.~\cite{CH13,IW14}
The electronic spin of a single NV center (hereafter NV center specifically refers to the NV$^-$ state) can be manipulated by microwave at around 2.87~GHz,
and be initialized and read out optically owing to the state-dependent fluorescence.~\cite{JW06}
Its long coherence time at room temperature makes it a leading candidate for a ultra-sensitive, nanoscale magnetometer.~\cite{BNT+09,CB05,D08,TCC+08,MSH+08,BCK+08}
In addition, the both long coherence time and magnetometry have also been realized in ensemble NVs under ambient conditions.~\cite{SPM+10,BPJ+13,ABL+09,PLS+11}
Consequently, NV-based sensing and imaging of magnetic fields, as well as other physical quantities such as temperatures and electric fields,
have become an active area of research and are expected to find practical applications
not only in physics but also in biology, medical science, planetary science, and so forth.~\cite{TDC+13,NJD+13,ABL+10,KMY+13,DFD+11,SCLD14,GLP+15,FWL+14}

In optically-detected magnetic resonance (ODMR) experiments of NV centers,
a thin straight copper wire or a millimeter-sized loop coil is commonly used to generate an oscillating magnetic field $\bm{B}_{\mathrm{ac}}$.~\cite{CDT+06,CTA+15}
For sensing with ensemble NVs, in which the sensitivity is enhanced by the square-root of the number of spins at the expense of nanoscale resolution,
the uniformity of $\bm{B}_{\mathrm{ac}}$ is particularly important.
Even in sensing with single NVs, the starting point is to search a diamond substrate for single NVs by confocal microscopy and ODMR.
$B_{\mathrm{ac}} = (|\bm{B}_{\mathrm{ac}}|)$ can be made strong in the vicinity of a wire, but decreases rapidly as departing from it.
On the other hand, $B_{\mathrm{ac}}$ generated by a loop coil is more uniform, but tends to be much weaker than that by the wire.
Moreover, the both structures are placed in contact with a diamond surface,
nearby which an objective lens and/or a specimen to be sensed, varying from inorganic materials to living cells, are necessarily present.
One cannot observe, for instance, the region beneath the wire, as it hinders the light propagation.
Therefore, it is preferable to leave the surface open as much as possible.

A few microwave circuits for ODMR of NV centers that are placed beneath diamond samples have been devised.
Bayat {\it et al.} developed a double split-ring resonator which achieves spatially uniform $B_{\mathrm{ac}}$ and efficient coupling to NV centers.~\cite{BCB+14}
While the resonator frequency is shown to be tunable within 400~MHz around 2.87~GHz by placing a copper tape near the structure,
its moderate quality factor ($Q \sim$ 70) allows only one resonance (typically $<$10~MHz linewidth) within the bandwidth (40~MHz),
making it difficult to simultaneously observe multiple resonances separated by tens to hundreds MHz under weak external magnetic fields.
Mr\'{o}zek {\it et al.} designed a microwave circuit operating between 2.7~GHz and 3.1~GHz
with a notable capability of generating arbitrary microwave polarizations.~\cite{MMRG15}
A drawback is that the distributions of microwave strength and polarization depend largely on the positions and careful calibrations may be necessary.

In this work, we aim to design and characterize a microwave planar ring antenna that addresses above-mentioned limitations.
The design targets of our antenna are as follows;
(i) It is placed beneath a diamond sample, thereby does not interfere with the objective lens or the specimen.
(ii) It generates $B_{\mathrm{ac}}$ spatially uniform within an area of about 1~mm$^2$, which covers large parts of typical diamond samples.
(iii) It has a resonance frequency at around 2.87~GHz and a bandwidth of a few 100~MHz ($Q <$~10) so that multiple resonances can be addressed.

\section{Design and S-parameter}
\begin{figure}
\begin{center}
\includegraphics{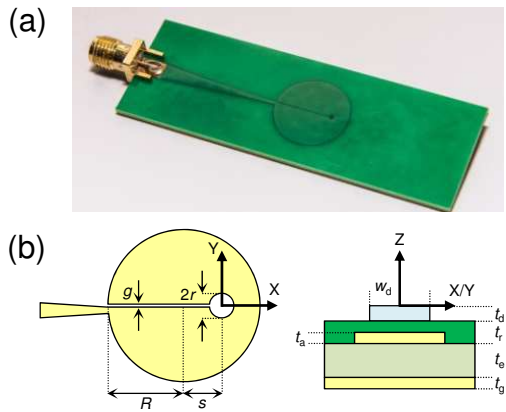}
\caption{(Color online)
(a) A photograph of a microwave antenna.
(b) Design parameters and the definition of the XYZ coordinate.
$R$ (= 7.0~mm): ring radius, $r$ (= 0.5~mm): hole radius, $s$ (= 3.9~mm): distance between the ring and hole centers, $g$ (= 0.1~mm): gap,
$w_\mathrm{d}$ (= 2.2~mm): width of diamond ($\epsilon_\mathrm{r}$ = 5.68, $\sigma$ = 0~S/m),
$t_\mathrm{d}$ (= 0.51~mm): thickness of diamond,
$t_\mathrm{r}$ (= 0.03~mm): thickness of resist (as a solder mask, $\epsilon_\mathrm{r}$ = 4.3, $\tan \delta$ = 0.025),
$t_{\mathrm{a}}$ (= 0.018~mm): thickness of metal conductor for antenna (copper, $\sigma$ = 5.8 $\times$ 10$^7$~S/m).
$t_e$ (= 1.6~mm): thickness of epoxy glass (FR4, $\epsilon_\mathrm{r}$ = 4.3, $\tan \delta$ = 0.03).
$t_{\mathrm{g}}$ (= 0~mm in simulation, 0.018~mm in fabricated devices): thickness of metal conductor for ground plane.
Here, $\epsilon_\mathrm{r}$ is the relative permittivity, $\sigma$ is the conductivity, and $\tan \delta$ is the loss tangent.
Simulation and actual values are the same except for $t_{\mathrm{g}}$.
To impedance-match with instrumental 50~$\Omega$, the width of the 23-mm-long feed line is consecutively varied from 3.3~mm (SMA connector side) to 0.54~mm (ring edge).
}
\label{fig1}
\end{center}
\end{figure}

Figure~\ref{fig1}(a) shows a photograph of a fabricated microwave antenna we have developed.
The design parameters are summarized in Fig.~\ref{fig1}(b).
Our ring resonator is regarded as a single-loop coil surrounding a circular hole,
inside of which resonant magnetic fields are concentrated.~\cite{BCB+14,HW81,ABG+05}
To help intuitive understanding of the design, we model the structure as a simple series LCR circuit.
The inductance $L_0$ arises from the circular hole: $L_0 \propto r$.
The capacitance $C_0$ is formed by the gap: $C_0 \propto (R+s-r)/g$.
The constriction connecting the upper and lower half circles will contribute to the resistance: $R_0 \propto (R-s-r)^{-1}$.
Then, the resonance frequency $2 \pi f_0 = (L_0 C_0)^{-1/2}$ and the quality factor $Q_0 = 2 \pi f_0 L_0/R_0$ are proportional to
$\sqrt{g/\{r(R+s-r)\}}$ and $(R-s-r)\sqrt{gr/(R+s-r)}$, respectively.
These expressions, although very naive, indicate that small $g$ and large $s$ are desirable to achieve low $Q$, which is our design target (iii).
Practically, the minimum achievable $g$ is limited by our manufacturing accuracy of 0.1~mm.

In relation to our design targets (i) and (ii), the primary consideration is the sample size.
Typical diamond substrates are 1 to 5~mm on a side with thickness $t_{\mathrm{d}}$ of around 0.5~mm.
As the NV centers existing near the top surface are used for sensing, we are mainly concerned with the distribution of $B_{\mathrm{ac}}$ about 0.5~mm above the antenna surface.
We find $r$ comparable with $t_{\mathrm{d}}$ is advantageous to deliver spatially uniform and strong $B_{\mathrm{ac}}$ to the diamond surface.
We thus limit $r$ to be around 0.5~mm, and adjust the remaining design parameters $R$ and $s$ to tune $f_0$ and $Q_0$ to the desired values,
using three-dimensional electromagnetic simulation software CST MICROWAVE STUDIO\textsuperscript{\textregistered}. 
In the simulation, we set the width and thickness of diamond as $w_\mathrm{d}$ = 2.2~mm and $t_\mathrm{d}$ = 0.51~mm, respectively,
to coincide with the actual size of the diamond sample used for $S$-parameter and ODMR measurements below.
However, the electromagnetic property of our antenna is found to be insensitive to the sizes of diamond samples.

\begin{figure}
\begin{center}
\includegraphics{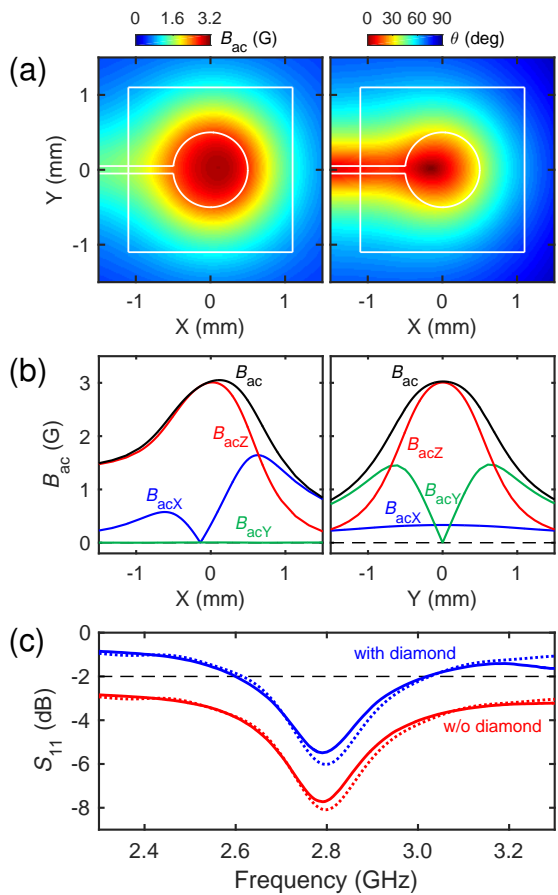}
\caption{(Color)
(a) Simulated $B_{\mathrm{ac}}$ and $\theta$ at the diamond surface (Z = 0~mm)
with $f_{\mathrm{mw}}$ = 2.87~GHz and $P_{\mathrm{mw}}$ = 1~W.
The solid (white) lines outline the diamond sample and the hole and gap structures of the antenna.
(b) Cross sections of (a) along the X (left) and Y (right) axes.
The components of $\bm{B}_{\mathrm{ac}}$ are also shown.
(c) Measured (solid lines) and simulated (dotted lines) $S_{11}$ with (blue) and without (red) diamond on top of the antenna.
The data without diamond are shifted downward by 2~dB for clarity.
}
\label{fig2}
\end{center}
\end{figure}

Simulated $B_{\mathrm{ac}}$ at the diamond surface is shown in the left-hand side of Fig.~\ref{fig2}(a),
and its cross sections along the X and Y axes are in Fig.~\ref{fig2}(b).
The microwave frequency $f_{\mathrm{mw}}$ is 2.87~GHz and the microwave power $P_{\mathrm{mw}}$ is 1~W.
As expected, $B_{\mathrm{ac}}$ is concentrated inside of the hole and the main component of $B_{\mathrm{ac}}$ is in the Z direction ($B_{\mathrm{acZ}} \geq$~0).
To evaluate the directional uniformity, we calculate the zenith angle $\theta$, {\it i.e.}, the tilt angle from the Z axis, as
\begin{equation}
\theta = \arccos \left( \frac{B_{\mathrm{acZ}}}{B_{\mathrm{ac}}} \right) \quad (0^{\circ} \leq \theta \leq 90^{\circ}),
\label{deftheta}
\end{equation}
which is shown in the right-hand side of Fig.~\ref{fig2}(a).
The X and Y components are very small ($\theta \sim$ 0$^{\circ}$) in the center region and increase towards the peripheral of the hole ($\theta \sim$ 30$^{\circ}$).
Thus, as long as we restrict the scanning range inside of the hole, we expect spatially and directionally uniform $B_{\mathrm{ac}}$.

Figure~\ref{fig2}(c) shows measured (using a vector network analyzer) and simulated reflection coefficients $S_{11}$ with and without diamond on top of the antenna.
The experimental resonance frequency is almost unaffected (2790~MHz) by the presence of diamond.
We obtain the bandwidth of 437~MHz ($Q \sim$ 6.4) and 395~MHz ($Q \sim$ 7) with and without diamond, respectively.
These features are well reproduced by simulation.~\cite{note1}

\section{Sample and experimental setup\label{sample}}
\begin{figure}
\begin{center}
\includegraphics{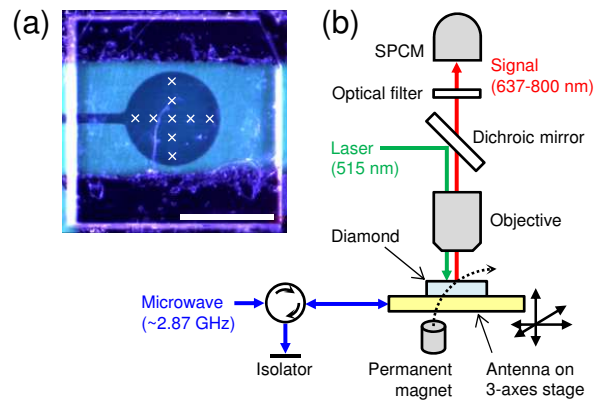}
\caption{(Color online)
(a) An optical microscope image of the diamond sample mounted on the antenna.
The scale bar is 1~mm.
The cross marks ($\times$) indicate the positions where ODMR measurements in Section~\ref{odmr} are conducted.
Their XY coordinates (in millimeter) are $(0, 0)$, $(0, \pm0.2)$, $(0, \pm0.4)$, $(\pm0.2, 0)$, and $(\pm0.4, 0)$.
(b) Schematic of experimental setup.
SPCM: single photon counting module.
}
\label{fig3}
\end{center}
\end{figure}

The ($w_\mathrm{d}^2 \times t_\mathrm{d}$)-sized diamond sample is pictured in Fig.~\ref{fig3}(a).
Through the transparent sample, the hole and gap structures of the antenna as well as double-sided tapes to immobilize the sample are visible.
Microwave plasma-assisted chemical vapor deposition is used to grow a 100-nm-thick N-doped diamond film on a type IIa (001) substrate.
A portion of doped N forms NV centers.~\cite{WKNS09,IFS+12,ORW+13}
The NV density is estimated to be of the order of 10$^{15}$~cm$^{-3}$.

The spin Hamiltonian of the NV ground state is given, in the frequency unit, as
\begin{equation}
\mathcal{H}^{(i)} = D (S_z^{(i)})^2 + \gamma \bm {B}_0 \cdot \bm{S}^{(i)}.
\label{spinhamiltonian}
\end{equation}
Here, $D$ = 2.87~GHz is the zero-field splitting,
$\gamma$ = 2.8~MHz/G is the electron magnetic moment, 
$\bm {B}_0$ is the external magnetic field felt by the NV centers, and
$\bm{S}^{(i)}$ is the $S$ = 1 spin operator with the quantization ($z$) axis taken as the $i$th NV axis.
The NV axis is the vector connecting N and V, and is parallel to one of four $\langle$111$\rangle$ crystallographic axes ($i$ = 1, 2, 3, 4).
In ODMR of ensemble NVs, the maximum of eight resonances, corresponding to the four possible NV axes and the two allowed $|\Delta m_S|$ = 1 transitions are observed.

Key elements of our experimental setup are shown in Fig.~\ref{fig3}(b).
A 515-nm green laser light non-resonantly excites the NV electrons,
and red signal photons emitted from the NV centers are counted with a single photon counting module.
The duration and interval of laser illumination are controlled by an acousto-optic modulator.

For continuous wave (CW) and pulsed magnetic resonance of the NV electronic spins, a microwave signal generator and an amplifier are connected to the antenna.
A permanent magnet is used to split the $m_S$ = $\pm$1 states.
In our measurements, the positions of the objective lens and the permanent magnet are fixed,
and the diamond sample and the antenna move in the XYZ directions with 3-axes piezo and translation stages.
Therefore, the orientation and strength of the external magnetic field do not change at the focal point.
This means that even when NV centers in different positions are measured they feel the same $\bm {B}_0$.

\section{CW and pulsed ODMR\label{odmr}}
\begin{figure}
\begin{center}
\includegraphics{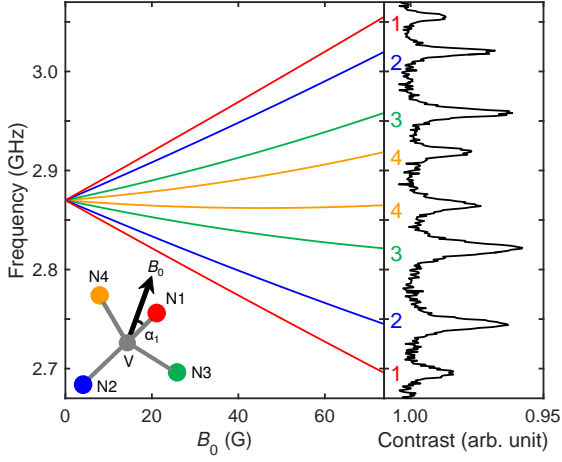}
\caption{(Color online)
CW ODMR spectrum taken at the hole center $(0, 0)$ (right) and calculated $B_0$-dependence of the transition frequencies at
$\alpha_1$ = 29$^{\circ}$, $\alpha_2$ = 132$^{\circ}$, $\alpha_3$ = 109$^{\circ}$, and $\alpha_4$ = 83$^{\circ}$ (left).
At $B_0$ = 74~G, the eight resonances are reproduced simultaneously.
}
\label{fig4}
\end{center}
\end{figure}

The CW ODMR spectrum taken at the hole center $(0, 0)$ is shown in the right-side of Fig.~\ref{fig4}.
In the CW mode, the laser illumination and the photon counting are executed continuously, during which the microwave frequency is swept.
The fluorescence decreases at the resonance, giving a series of dips.
The contrast is given as the ratio between the data with the microwave on and off in order to cancel out the background.
The eight dips characteristic of ensemble NVs are clearly observed.
The Zeeman term of Eq.~(\ref{spinhamiltonian}) is written as $\gamma B_0 (\sin \alpha_i S_x^{(i)} + \cos \alpha_i S_z^{(i)})$, where $\alpha_i$ is defined as the angle between $\bm{B}_0$ and the $i$th NV axis.
By numerically solving Eq.~(\ref{spinhamiltonian}), we determine $B_0$ and $\alpha_i$ as
$B_0$ = 74~G with $\alpha_1$ = 29$^{\circ}$, $\alpha_2$ = 132$^{\circ}$, $\alpha_3$ = 109$^{\circ}$, and $\alpha_4$ = 83$^{\circ}$
that reproduce the eight resonances simultaneously.
As $\alpha_i$ is made closer to 90$^{\circ}$, the off-diagonal term with $S_x^{(i)}$ becomes non-negligible and the nonlinearity in the evolution of transition frequencies is more pronounced.

The observations here demonstrate that under the external magnetic field up to 100~G or so,
which is a typical condition for NV-based sensing and imaging experiments,~\cite{CTA+15,MWF+10}
our antenna does not require an adjustment of its resonance frequency by external perturbations (such as placing a copper tape as in the case of Ref.~\onlinecite{BCB+14}).

\begin{figure}
\begin{center}
\includegraphics{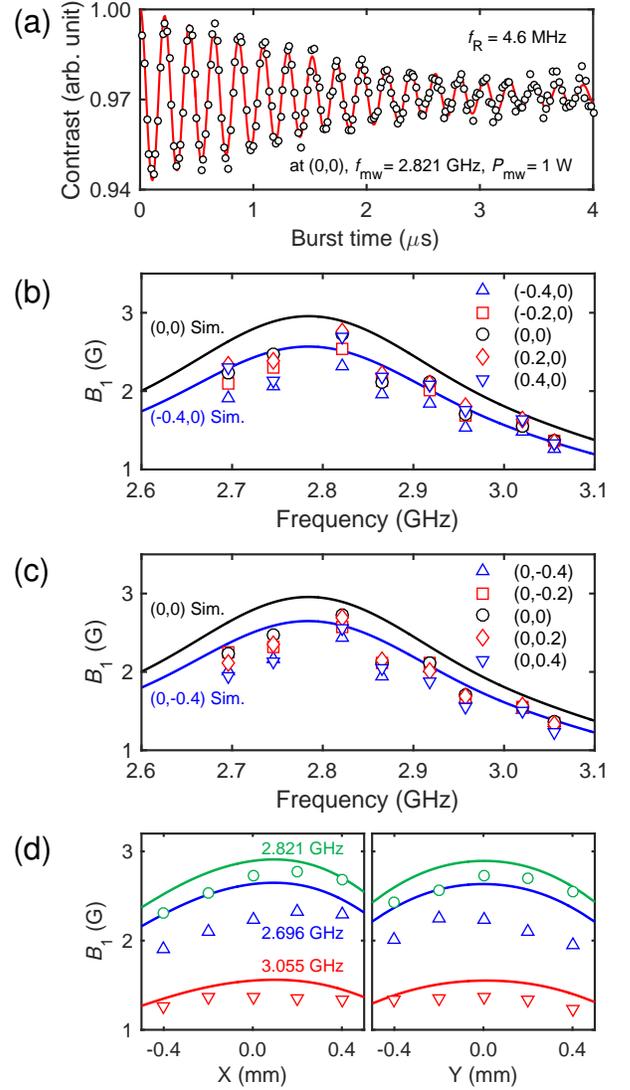}
\caption{(Color online)
(a) An example of Rabi oscillations.
The solid line is the fit.
(b, c) Plots of $B_1$ at $P_{\mathrm{mw}}$ = 1~W for five positions along the X (b) and Y (c) axes.
The solid lines are simulation results at $(0, 0)$, $(-0.4, 0)$ and $(-0.4, 0)$ as indicated.
(d) Experimental and simulated (solid lines) spatial distributions of $B_1$ along the X (left) and Y (right) axes at three frequencies.
}
\label{fig5}
\end{center}
\end{figure}

Next, we carry out pulsed ODMR to observe Rabi oscillations and examine the oscillating magnetic field generated by the antenna.
After initialization into the $m_S$ = 0 state by the laser illumination for 5~$\mu$s, microwave is burst for a short time.
The spin state is read out by the second laser illumination.
As the burst time $t$ is varied, the contrast oscillates reflecting the occupation of the $m_S$ = 0 state.
Figure~\ref{fig5}(a) shows an example of measured Rabi oscillations.
The data is fitted by the functional form $A \exp(-t/\tau_\mathrm{R}) \cos(2 \pi f_\mathrm{R} t)+B$,
where $A$ is the amplitude, $B$ is the offset, $\tau_\mathrm{R}$ is the decay timescale of the Rabi oscillation, and $f_\mathrm{R}$ is the Rabi frequency.

We define $\bm{B}_1$ as $\bm{B}_{\mathrm{ac}}$ generated at the focal point ({\it i.e.,} the measurement position).
At given $P_{\mathrm{mw}}$, $f_{\mathrm{R}}$ and $B_1$ are related by
\begin{equation}
f_{\mathrm{R}} = \frac{\gamma B_1 \sin \beta_i}{\sqrt{2}},
\label{frabi}
\end{equation}
where $\beta_i$ is the angle between $\bm{B}_1$ and the $i$th NV axis.
The components of $\bm{B}_1$ perpendicular to the NV axis rotates the NV electronic spin.
The factor $1/\sqrt{2}$ is due to the rotating wave approximation in the $S$ = 1 spin system.~\cite{AJ01,HDF+08}
Through the CW ODMR measurements we already know the NV axes responsible for the respective resonances [Fig.\ref{fig4}],
and the simulation results give the orientation of $\bm{B}_1$ [Fig.\ref{fig2}(a)].
They together give $\beta_i$, allowing us to convert experimental $f_{\mathrm{R}}$ into $B_1$.

Figures~\ref{fig5}(b) and (c) plot in total 72 values of $B_1$ obtained at nine positions [Fig.~\ref{fig3}(a)] and eight frequencies [Fig.\ref{fig4}], covering wide spatial and frequencies ranges.
There, $B_1$ deduced solely from simulation are also shown for $(0, 0)$ and $(0, -0.4)$ or $(-0.4, 0)$.
Other simulated $B_1$ lines lie between these lines and are not shown.
In this simulation, the presence of the double-sided tapes, the thickness of which is 70~$\mu$m, are taken into account.
This results in a reduction of $B_1$ compared with Fig.~\ref{fig2}.
Spatial distributions of $B_1$ along the X and Y axes for three representative frequencies are also shown in Fig.~\ref{fig5}(d).
Overall, the experiments and simulation show similar behaviors with the experimental values slightly smaller than the simulated values.
Based on these results, we conclude that our antenna achieves high spatial uniformity and high bandwidth at the same time.

As a remark, the fastest $f_{\mathrm{R}}$ achieved in our current setup is 10~MHz, which is limited by the saturation power of the microwave amplifier we use.
In this condition, we have been able to conduct multi-pulse experiments (dynamical decoupling with hundreds of pulses) without any heating problems. 
The constraint set by the antenna itself is its bandwidth of 400~MHz, with which the shortest pulse length of 2.5~ns should in principle be possible.

\section{Comparison with other methods}
\begin{figure}
\begin{center}
\includegraphics{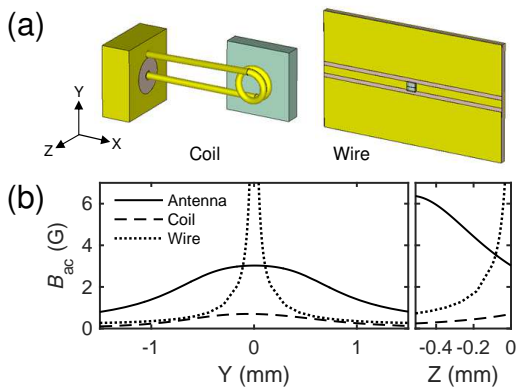}
\caption{(Color online)
Simulation and comparison of different methods for applying oscillating magnetic fields.
(a) CAD designs of coil- and wire-based microwave setups.
The plates (in moss green) represent diamond substrates with dimensions $w_\mathrm{d}^2 \times t_\mathrm{d}$~mm$^3$.
On the left is a 1.5-turn coil with the wire diameter of 220~$\mu$m and the loop diameter of 1~mm placed in contact with the diamond substrate.
One end of the wire is connected to the inner conductor of a 50~$\Omega$ connector, through which microwave is supplied.
The other end of the wire is terminated at the grounded outer conductor.
On the right is a coplanar waveguide matched to 50~$\Omega$ with the width of the center line of 2~mm and the gaps of 1~mm.
The center line is removed beneath diamond, and instead a 30-$\mu$m-diameter copper wire bridges the two ends of the center line.
(b) Distribution of $B_\mathrm{ac}$ along the Y (left) and Z (right) axes at $f_{\mathrm{mw}}$ = 2.87~GHz and $P_{\mathrm{mw}}$ = 1~W.
}
\label{fig6}
\end{center}
\end{figure}

Finally, we make a brief comparison between different methods for applying oscillating magnetic fields to diamond.
We simulate coil- and wire-based microwave setups as modeled in Fig.~\ref{fig6}(a).
Figure~\ref{fig6}(b) shows the distribution of $B_{\mathrm{ac}}$ along the Y and Z axes.
In all models, the origin of the coordinate is set at the center of the diamond surface.
As mentioned in Section~\ref{intro}, $B_{\mathrm{ac}}$ is strong near the wire, but weakens as the inverse of the distance from the wire (Amp\`{e}re's law).
The coil produces more or less uniform $B_{\mathrm{ac}}$ inside of the loop, but the strength is much weaker.
The antenna combines the both features.
It is also worth noting that $B_{\mathrm{ac}}$ of the antenna becomes stronger inside of the diamond.
This is advantageous when NV centers deep inside the substrate are investigated, which is important for materials characterization as well as for quantum information processing.
When the NV electronic spins are used as quantum bits, the depth from the substrate may be less critical than in sensing application,
and the deep NV centers may be preferred for their stability and longer coherence times.

\section{Conclusion}
In conclusion, we have designed and characterized a microwave planar ring antenna suitable for ODMR of NV centers in diamond.
It has the resonance frequency at around 2.87~GHz with the bandwidth of 400~MHz,
ensuring that ODMR can be observed under external magnetic fields up to 100~G without the need of adjustment of the resonance frequency.
It is also spatially uniform within the 1-mm-diameter hole, enabling the magnetic-field imaging in the wide spatial range.
For imaging experiments, with a piezoelectric actuators one can scan a few 100$^2$~$\mu$m$^2$ areas and with a galvanometric mirror even larger areas can be scanned.
NV-based wide-field imaging techniques using a CCD or CMOS camera are also compatible with the antenna.~\cite{PLS+11,GLP+15} 
These features will facilitate the experiments on quantum sensing and imaging using NV centers at room temperatures.

\section*{Acknowledgments}
HW acknowledges the support from JSPS Grant-in-Aid for Scientific Research (KAKENHI) (A) No.~26249108 and
JST Development of Systems and Technologies for Advanced Measurement and Analysis (SENTAN).
JI-H acknowledges the support from KAKENHI (B) No.~15H03996 and on Innovative Areas No.~15H05868, and Advanced Photon Science Alliance.
KMI acknowledges the support from KAKENHI (S) No.~26220602 and JSPS Core-to-Core Program.


\begin{thebibliography}{34}
\bibitem [{not()}]{note1}%
  \BibitemOpen
  \href@noop {} {}\bibinfo {note} {We note that in this simulation the presence
  of double-sided tapes [Fig.~\ref{fig3}(a)] is not considered. Also, in
  simulation the antenna metal is ``embedded'' in the resist layer, and the
  thickness of the resist changes depending on the presence or absence of the
  metal [Fig.~\ref{fig1}(b)]. In actual fabricated antennas, the resist covers
  the surfaces of metal and epoxy glass with uniform thickness.
  Figure~\ref{fig2}(c) confirms these differences have minor, if not
  negligible, effects on comparison between simulation and
  measurement.}\BibitemShut {Stop}%
\end{thebibliography}
\end{document}